\documentclass[aps,prd,amssymb,cite,
amsfonts,epsf,preprintnumbers,nofootinbib,superscriptaddress]{revtex4}

\usepackage[dvips]{graphicx}
\usepackage{bm,latexsym,amsmath,amssymb,amsfonts}
\usepackage[usenames,dvipsnames]{color}
\usepackage[colorlinks=true,linkcolor=blue]{hyperref}
\usepackage{color}
\usepackage{soul}
\usepackage{epsfig}
\usepackage{braket}	
\usepackage[mathscr]{eucal}
\usepackage{cancel}
\usepackage{mathrsfs}
\usepackage{pgf, tikz}
\usepackage{slashed}
\usetikzlibrary{arrows,automata}
 
\definecolor{mypink1}{rgb}{0.858, 0.188, 0.478}
\definecolor{mypink2}{RGB}{219, 48, 122}
\definecolor{mypink3}{cmyk}{0, 0.7808, 0.4429, 0.1412}
\definecolor{mygray}{gray}{0.6}
\definecolor{pptbg}{rgb}{0.961,0.945,0.863}

\newcommand{\be}[1]{\begin{equation} \label{#1}}
\newcommand{\ee}{\end{equation}}
\newcommand{\bea}{\begin{eqnarray}}
\newcommand{\eea}{\end{eqnarray}}
\newcommand{\ba}{\begin{array}}
\newcommand{\ea}{\end{array}}
\newcommand{\nn}{\nonumber}
\newcommand{\tcb}{\textcolor{blue}}

\newcommand{\cE}{\cS}

\newcommand{\cS}{\mathscr{S}}

\newcommand{\nA}{\slashed{\mathcal{A}}}

\newcommand{\bel}{\begin{align}}
\newcommand{\eel}{\end{align}}

\begin{document}
\title{Nonadiabaticity of Quantum harmonic oscillators}

\author{Hyeong-Chan Kim and Youngone Lee}
\affiliation{School of Liberal Arts and Sciences, Korea National University of Transportation, Chungju 380-702, Republic of Korea}

\email{hckim@ut.ac.kr,  youngone@ut.ac.kr}

\begin{abstract}
We propose a quantity, $\nA$, as a measure describing the nonadiabaticity of a thermodynamic process.
For this purpose, we use a schematic method to find the measure of the `degree of nonadiabaticity'.
The method utilizes an `invariant' thermal state constructed from the Ermakov-Lewis-Riesenfeld invariant.
Specifically, we study a frequency-modulated quantum harmonic oscillator as a thermodynamic system.
Naturally, we write the first law of thermodynamics with $\nA$ as a measurable quantity.
We discuss universality for the method and some possible applications.
\end{abstract}
\maketitle

\section{Introduction}

In the study of non-equilibrium thermodynamics, we focus mainly on 
temporal changes of global statistical quantities, e.g., heat, temperature, et cetera.
In principle, any thermodynamical system can be described by a sum of quantum fields, which are described by a weighted sum of oscillators of different frequencies. 
In addition, one can approximate a local minimum of a potential utilizing a harmonic oscillator potential.
Actually, the oscillators have been studied frequently in thermodynamics.
Davies~\cite{davies1973} show that a quantum harmonic oscillator in an infinite heat bath relaxes to a canonical thermal state.
Rezek~\cite{phdthesis} studied the thermodynamic engine based on working fluid composed of quantum harmonic oscillators.
Deffner and Lutz~\cite{Deffner_2008} examined the quantum harmonic oscillator with arbitrary frequency modulation from a thermodynamic point of view.
Recently, by using the thermodynamics of a classical harmonic oscillator, Boyer~\cite{Boyer:2019irq}
discussed that the third law of thermodynamics leads to the Planck spectrum without resorting to quantum theory.
Therefore, as a probe, quantum harmonic oscillators can be the first system to study.

Recently, there are many studies about qubits coupled to harmonic oscillators~\cite{Qubitosc,brunelli2012qubit,Pechal}.
In quantum computing, decoherence by thermalization causes errors in calculations~\cite{errorcorrection}.
Therefore, it is important to know how thermalization happens in an oscillator system.
The process is generally nonadiabatic in nature.
It would be interesting to find an analytical method which deals the non-adiabatic natures without resorting to any numerical technique. 
It is even more beneficial if 
 the method gives us a unified viewpoint for time-varying thermodynamic systems other than a simple oscillator.
In this article, we present such a method for a nonadiabatic process as a first step towards more general analysis including thermalization.
We also show that the method gives rise to a characteristic quantity to which we will assign  {\it `nonadiabaticity'} $\nA$.
Thus an adiabatic process naturally belongs to the class of processes that the value of $\nA$ is negligible.

In this paper, we study 
a frequency-modulated harmonic oscillator with the hamiltonian
\be{H:HO2}
\hat H(t)= \frac{\hat p^2}{2m} + \frac12 m \omega^2(t) \hat x^2,
\ee
where the mass $m$ is a constant but the frequency $\omega$ changes with time.
From the form of the frequency-modulated hamiltonian above, one can say that the  process is adiabatic when the frequency is slowly changing or quasi-static,  nonadiabatic otherwise.
In other words, a dimensionless quantity measuring temporal changes is small enough, $\dot\omega/\omega^2\approx 0$.
We need to keep in mind that this terminology in quantum mechanics is to be distinguished from the traditional term `adiabatic' in thermodynamics denoting the absence of heat transfer.

Our assumption here is that we can approximate the thermal quantum state of the oscillator with a Gaussian-type density matrix~\cite{Polkovnikov:2010yn,Calabrese:2011vdk}.
That is, the reduced density matrix of a subsystem belonging to an infinite system was shown to be described in terms of Gibbs distribution or generalized Gibbs ensemble.

We first review the  Ermakov-Lewis-Riesenfeld (ELR) invariant for frequency-modulated quantum harmonic oscillators in Sec.~\ref{review}.
We define a key quantity describing `squeezing energy' at the end of the section.
In Sec.~\ref{example}, we present an example of a frequency-modulated quantum harmonic oscillator
undergoing nonadiabatic changes.
We calculate the behavior of the oscillator and the `squeezing energy'. 
In Sec.~\ref{na-thermo}, we study thermodynamics of a frequency-modulated quantum harmonic oscillator.
We quantify nonadiabaticity of a quantum state. 
We also write down the first law of thermodynamics with the `nonadiabaticity'  $\nA$.
Lastly, we discuss the universality of our `invariant method'
to broaden the applicability beyond the harmonic oscillator in Summary~\ref{summary}.

\section{ Review of The Ermakov-Lewis-Riesenfeld invariant for quantum harmonic oscillators}\label{review}
 
\subsection{Review of the ELR invariant}

Knowledges on constants of motion or invariants of a system usually simplify analysis and present insights.
The ELR invariant was introduced
by Ermakov\cite{Ermakov}, Lewis, and Riesenfeld~\cite{PhysRevLett.18.510,doi:10.1063/1.1664991}
for solving a time-dependent quantum system.
For a Hamiltonian operator $\hat H(t)$, they assumed that there exists a non-trivial Hermitian operator $\hat I(t)$ which does not change with time. 
Thus, $\hat I(t)$ satisfies 
$$
\frac{d\hat I}{dt}\equiv \frac{\partial \hat I}{\partial t}+i[\hat H,\hat I]=0.
$$
The invariant they found for a quantum harmonic oscillator with the hamiltonian~\eqref{H:HO2} takes the form, 
\be{I2}
\hat I(t)
=g_+(t)\frac{m \hat x^2}{2}+g_-(t) \frac{\hat p^2}{2m} + g_0(t) \frac{\hat p \hat x + \hat x \hat p}{2},
\ee
where the temporal functions $g_0(t),g_\pm(t)$ satisfy certain mutual relations.
The authors also presented another invariant for a system of charged particle in a time-dependent electromagnetic field.

The method of using the ELR invariant was also used in quantum computing to develop a quick quantum algorithms running on a nonadiabatic regime~\cite{SARANDY20113343,Chen_2011}.
The invariant above is a quadratic function of the position and the momentum.
Later in Refs.~\cite{Kim:1996vm,Lee:1997rb,Ji:1998dh}, the authors found a simple way to obtain the invariant creation/annihilation operator directly.
This method is appropriate for this work. Thus we briefly summarize it here.

First, we obtain an invariant annihilation operator $\hat b$ and its conjugate $\hat b^\dagger$ as:
\be{LR eq}
\frac{d \hat b}{dt} \equiv  \frac{\partial \hat b}{\partial t} +i [\hat H, \hat b] =0.
\ee
We impose the normalization condition $[\hat b, \hat b^\dagger] =1$.
It is enough to choose $\hat b$ and $\hat b^\dagger$ as linear combinations of $\hat x$ and $\hat p$.
It is because the Hamiltonian is a quadratic function of them.
Higher-order invariants can be constructed by the products of the linear ones.
Especially, the quadratic invariant $\hat I$ may take the form: $\hat I\sim \hat b^\dagger \hat b+1/2$.

We can find an invariant annihilation operator as a linear combination of the position and the momentum operators:
\be{b}
\hat b = f(t) \hat x(t) +i g(t) \hat p(t),
\ee
where $f$ and $g$ are complex functions. 
Equation~\eqref{LR eq} gives two linear differential equation for $f$ and $g$, 
\be{fg:de}
f =-i m\dot g,~~\dot f=im \omega^2(t) g~\Rightarrow~\ddot g + \omega^2(t) g =0.
\ee
Note that $g$ satisfies the classical equation of motion of a harmonic oscillator with time-dependent frequency $\omega(t)$.
In this sense, the full quantum evolutions of a harmonic oscillator are given once we know a classical solution of the oscillator.
This is far simpler than performing the operator product calculation of the unitary evolution.
In addition, the function is closely related to the time-evolution of the temperature for a thermal state as will be shown later in this work.

The dimensionless function $g_-(t)$ is the square root of the non-oscillatory part of the classical solution $g$.
Then, one obtains the following:
\be{fg}
f = \left(\sqrt{\frac{m\omega_I}{2 g_-}}
	- \frac{i \dot g_-}{2} \sqrt{\frac{m}{2 \omega_I g_-}}\right)
		 e^{i \Theta(t)} , \qquad
g =  \sqrt{\frac{g_-}{2m \omega_I}} e^{i \Theta(t)} , \qquad
\Theta(t) \equiv \omega_I\int_{t_0}^t \frac{d\tau}{g_-(\tau)} ,
\ee
where $\omega_I$  denotes an (arbitrary) initial frequency.
Putting $f$ and $g$ into Eq.~\eqref{b}, the invariant annihilation operator $\hat b$ becomes
\be{invariant op}
\hat b \equiv e^{i \Theta(t)} \hat b(t), \qquad
\hat b(t) \equiv \left(1
	-\frac{ i \dot g_-}{2 \omega_I }
		 \right) \sqrt{\frac{m\omega_I}{2 g_-}} \hat x(t) + i \sqrt{\frac{g_-}{2m  \omega_I}} \hat p(t).
\ee 
Note that $\hat b(t)$ and $\hat b^\dagger(t)$ are not invariant while $\hat b$ and $\hat b^\dagger$ are invariant.

The quadratic invariant $\hat I(t)$ can be constructed by using the $\hat b$ and $\hat b^\dagger$ to be
\be{I}
\hat I(t) =  \omega_I \Big[\hat b^\dagger(t) \hat b(t) + \frac12 \Big].
\ee
Putting Eq.~\eqref{invariant op} into Eq.~\eqref{I}, we can write the invariant into the ELR form~\cite{PhysRevLett.18.510}.
Matching  equations~\eqref{I2} and \eqref{I} gives the relations between $g_\pm$ and $g_0$:
\be{g0g+}
g_0=-\frac{\dot g_-}{2},~~g_+=\frac{\omega_I^2}{g_-}\left(1+\frac{\dot g_-^2}{4 \omega_I^2} \right),
\ee
with a differential equation for $g_-$, which has a central importance in defining the nonadiabaticity,
\be{ddot g-}
\frac{1}{2}\frac{\ddot g_-}{ g_-} =\frac{1}{4}\frac{\dot g_-^2}{g_-^2} 
+\omega_I^2\left(\frac{1}{g_-^2}-\frac{\omega^2}{\omega_I^2}\right).
\ee
If we adjust constants of motion so that $\hat I=\hat H_0$, where $\hat H_0$ denotes the initial hamiltonian at $t=t_0$ and $\omega_I=\omega(t_0)$.
Once we know $g_-$ and $\dot g_-$ at an initial time, the whole evolution of the oscillator is determined from this equation.
In other words, once we know $g_-$ and $\dot g_-$ at a given time, the quantum Fock space at $t$ can be constructed from the eigenstates of $\hat I$.

The invariant $\hat I(t)$ in~\eqref{I2} differs from the hamiltonian in \eqref{H:HO2} by
\be{I:H g}
\hat I(t) - g_- \hat H = \frac{m \ddot g_-}{4} \hat x^2 - \frac{\dot g_-}{4} (\hat x \hat p+\hat p \hat x).
\ee
For a quasi-static process, $\dot g_-,\ddot g_-\sim0$ and $\hat I\sim g_- \hat H$.
Therefore, one can always write down $\hat I(t)$ in a $g_-$ scaled basis of $\hat H(t)$ for such a process.
The frequency $\omega_I$ can be written in terms of $g_\pm$ and $g_0$, to be consistent,
\be{omega I}
 \omega_I \equiv \sqrt{g_+g_--g_0^2} .
\ee

Let us discuss how $g_-$ behaves.
Defining $h\equiv \sqrt{g_-}$ (with $g_->0$) rewrites the equation~\eqref{ddot g-} as
\be{eomh}
\ddot h=-\omega^2(t) h+\frac{\omega_I^2}{h^3}.
\ee
This is the classical equation of motion of a particle under two external forces,
 $F_1=-\omega(t)^2 h$ and $F_2=\omega_I^2/h^3$.
Unlike $F_2$, $F_1$ is not a central force due to the time dependence.
When $\omega(t)$ is slowly varying, the force $F_1$ can be approximated as a central force.
In this case, we know that there is an energy-like quantity $\cS_{\rm adiabatic}$, which is conserved approximately.
In general, the energy of the system changes because of the time dependence of $\omega(t)$ respecting the external work done by the force $F_1$.
One can obtain such a quantity, $\cS(t)$, considering the contributions of the two forces, $F_1+F_2$,
that generalize $\cS_{\rm adiabatic}$.
By multiplying $\dot h$ on both sides of equal sign in Eq.~\eqref{eomh} and integrating by parts
we obtain the quantity
\be{eom E}
\cE(t) =\frac{\dot h^2}{2 \omega_I^2} +V(\omega, h) ,
\ee
where the potential-like term is
\be{E V}
V(\omega, h) \equiv \frac{1}2 \left(\frac{1}{h}
	-\frac{\omega(t)}{\omega_I} h \right)^2.
\ee
The explicit form of this dimensionless quantity\footnote{The quantity $\cS$ is dimensionless. To recover the energy dimension, we can multiply $\omega_I$ without loss of generality because it is invariant under the time evolution.} $\cS$ is defined by
\be{non-adia}
\qquad
\cS(t) \equiv \cS_{0}+
\frac{1}{2 \omega_I^2}
	\int_{t_0}^{t} \Big( h^2(t')- \frac{\omega_I}{\omega} \Big)\frac{d\omega^2(t')}{dt'}  dt' ,
\ee
where we choose $\cE_0$ is the corresponding value at the initial time $t_0$.
We choose $\cE =0 $ when the invariant is proportional to the hamiltonian.
Therefore, if we choose the invariant to be the same as the Hamiltonian at $t_0$, we can safely set $\cE_0 =0$.
Note that this energy-like term $\cS$ is always non-negative  
because it consists of  two perfect squares.
As we see later, this quantity $\cE$ is related to the `squeezing energy' for the quantum oscillator.
Later in this work, we call $\cE$ `squeezing factor' or $\cS$-factor. 

\subsection{Time evolution of the operators and Density matrix}

Rewriting Eq.~\eqref{invariant op}, we get the time-evolution of the position $\hat x(t)$ and the momentum $\hat p(t)$:
\be{x p:t}
\hat x(t) = \sqrt{\frac{g_-}{2m\omega_I}} [\hat b(t) + \hat b^\dagger(t)], \qquad
\hat p(t) = \sqrt{\frac{m  \omega_I}{2g_-}} \left\{
	i[\hat b^\dagger(t) -\hat b(t)]
	+\frac{\dot g_-}{2 \omega_I}
	[ \hat b(t) + \hat b^\dagger(t)]  \right\}.
\ee
Now, the time evolution of an operator $\hat A(t)$ can be interpreted as:
$$
\hat A_{\rm Sch}\equiv \hat A_{\rm Sch}\big(t,\hat x,\hat p\big)
~~\rightarrow~~
\hat A(t)=\hat A\big(t, \hat x(t),\hat p(t)\big),
$$
where the subscript ${}_{\rm Sch}$ denotes that the operator follows the Schr\"{o}dinger picture.
For example, the hamiltonian following this prescription can be written as
\be{H t>0}
\hat H 
= -\frac{ \omega_I}{4g_-} (\hat b^\dagger(t) -\hat b(t) )^2
	+\frac{i \dot g_-}{4g_-} (\hat b^{\dagger 2}(t) - \hat b^2(t)) +
	\left(\frac{ \dot g_-^2}{16\omega_I g_-} + \frac{g_-(t)\omega^2(t)}{4\omega_I }\right) (\hat b(t)+ \hat b^\dagger(t) )^2 .
\ee

Since the density matrix $\hat\rho(t)$ transforms as a state in the Schr{\"o}dinger picture,
the expectation value of an operator $\hat A$ at time $t$  is given by,
\bea
\overline{\braket{\hat A}}_{\rm Sch} (t)
\equiv {\rm Tr}\left[\hat\rho(t) \hat A_{\rm Sch}\right]
={\rm Tr}\left[U(t) \hat\rho_0U^\dagger(t) \hat A_{\rm Sch}\right],
~~~~
 U(t) =\lim_{N\to \infty} \prod_{k}e^{-i \hat H \left(\frac kN t\right)\frac tN},
\eea
where $\hat\rho_0$ is the initial state.
This expectation value is the same as that in the Heisenberg picture,
\bea\label{Heigen}
\overline{\braket{\hat A(t)}}_{\rm Hei}=
{\rm Tr}\left[\hat\rho_0 \hat A(t)\right],
~~~\hat A(t)\equiv U^\dagger(t)\hat  A(0)U(t).
\eea
As seen here, we drop the subscript ${}_{\rm Hei}$ for Heisenberg operators.

Since $\hat\rho_0$ is a constant matrix defined at the initial time, replacing it with an invariant matrix $\hat\rho_I(t)$ satisfying $\hat\rho_I(t)=\hat\rho_0$ does not alter the physics of the system.
We suggest an ELR invariant state  $\hat\rho_I(t)$ by using the ELR invariant $\hat I(t)$ as follows:
\be{rho 0}
\hat\rho_0=\rho_c e^{-\hat H_0/T_0}~\rightarrow~\hat\rho_I(t)=\rho_c e^{-\hat I(t)/T_0},
\ee
where $T_0$ and $\rho_c$ denote the initial temperature and the normalization constant of the state, respectively.

For a frequency-modulated quantum harmonic oscillator, 
a function $g_-(t)$ is sufficient to determine the quadratic invariant $\hat I(t)$ or the linear invariant annihilation operator $\hat b$.
Hence, 
the use of the invariant state gives clear advantages over traditional analysis
for studying nonadiabatic processes because the knowledge of the scalar function $g_-(t)$ determines the whole time evolution of all operators.

\section{ Squeezing factor for a frequency-modulated oscillator}\label{example}

Non-vanishing $\cS$-factor, $\cS \neq 0$,
implies that the mode solution $g$  in Eq.~\eqref{fg} fails to be a pure positive mode but a mixture of the positive and the negative modes. 
Let us construct the invariant vacuum $ |0\rangle_I$ and the corresponding quantum states
\be{I:n}
\hat b |0\rangle_I =0, \qquad |n\rangle_I =\frac1{\sqrt{n!}} (\hat b^\dagger)^n |0\rangle_I.
\ee
Then, the ground state $|0\rangle_I$ of the ELR invariant $\hat I$ is nothing but a squeezed state of the ground state of the hamiltonian~\cite{Kim:2003mi} with the squeezing parameter $r$.
The $\cS$-factor is related to the  squeezing parameter $r$  as:
\be{r}
\cS = \frac{2\omega}{\omega_I} \sinh^2r .
\ee
Therefore, $\cS$ vanishes when $r=0$. 
However, the explicit value of $\cS$ for a given $r$ changes with the frequency, representing the energy required to squeeze the state.
 
Now, let us consider an explicit example which allows exact analytic solutions with the frequency
\be{omega t}
\omega^2(t) = \omega_0^2-\varepsilon^2(t), \qquad
\varepsilon(t) = \varepsilon_- \frac{1-\tanh t/d }{2} + \varepsilon_+ \frac{1+\tanh t/d } {2} .
\ee
The frequency varies from $\omega_- = \sqrt{\omega_0^2 - \varepsilon_-^2}$ to $\omega_+ = \sqrt{\omega_0^2-
\varepsilon_+^2}$.
The exact solution to Eq.~\eqref{fg:de} is given by
\bea
g(t) &=& \frac1{\sqrt{2m\omega_I} }e^{-i (\omega_++\omega_-)\frac{t}2 }
\Big(\cosh \frac{t}{d} \Big)^{-i (\omega_+-\omega_-)\frac{d}{2}}
{}_2F_1(\alpha_-,\alpha_+;1-i \omega_- d; y), \label{sol:exact}
\eea
where ${}_2F_1$ denotes the Gaussian hypergeometric function and
\bea
y\equiv \frac{1+\tanh \frac{t}d }{2}, \qquad
\alpha_\pm = \frac{1\pm \sqrt{1+(\varepsilon_+-\varepsilon_-)^2 d^2}}{2} +i \frac{(\omega_+-\omega_-)d}{2} .
\eea
In the past infinity, we choose $g(t)$ in Eq.~\eqref{sol:exact} to be a pure positive frequency mode with $\omega= \omega_-$.
Then, the solution in the limit becomes 
\be{t neg}
\lim_{t \to -\infty} g(t)=\frac{2^{i(\omega_+-\omega_-)d/2} }{\sqrt{2m\omega_I} }e^{-i \omega_- t},  \qquad  \omega_I = \omega_- , \qquad g_- = 1 , \qquad \dot g_- =0.
\ee
The $\cS$-factor at that time becomes $\cE_i =0$.
At the future infinity, it generally becomes a mixture of the positive and the negative modes:
\be{t inf}
\lim_{t \to \infty}g(t) = \frac{2^{i(\omega_+-\omega_-)d/2}}{\sqrt{2m\omega_I} }  \left( \alpha e^{-i \omega_+ t} + \beta e^{i\omega_+ t}\right) ,
\ee
where
\bea
\alpha =  \frac{\Gamma(1-i\omega_- d) \Gamma(1-i \omega_- d-\alpha_- -\alpha_+)}{\Gamma(1-i\omega_- d -\alpha_-) \Gamma(1-i\omega_- d- \alpha_+)},  \qquad
\beta =  \frac{\Gamma(1-i\omega_- d) \Gamma(\alpha_- +\alpha_+-1+i \omega_- d)}{\Gamma(\alpha_-) \Gamma( \alpha_+)}.
\eea
Here $\Gamma$ denotes the gamma function.
For later applications, we show the absolute squares of $\alpha$ and $\beta$,
\bea
|\alpha|^2 = \frac12 \frac{\omega_-}{\omega_+} \frac{\cosh\pi (\omega_++\omega_-)d + \cos 2\pi x}{\sinh \pi \omega_- d \sinh \pi \omega_+ d}, \qquad
|\beta|^2 = \frac12 \frac{\omega_-}{\omega_+}
	\frac{\cosh\pi (\omega_+-\omega_-)d
	+ \cos 2\pi x}{\sinh \pi \omega_- d \sinh \pi \omega_+ d},
\eea
where $x = \sqrt{1+ (\varepsilon_+-\varepsilon_-)^2d^2}/2$.
The functions $\alpha$ and $\beta$ satisfy a Bogoliubov-type relation
$$
|\alpha|^2-|\beta|^2 = \frac{\omega_+}{\omega_-}.
$$

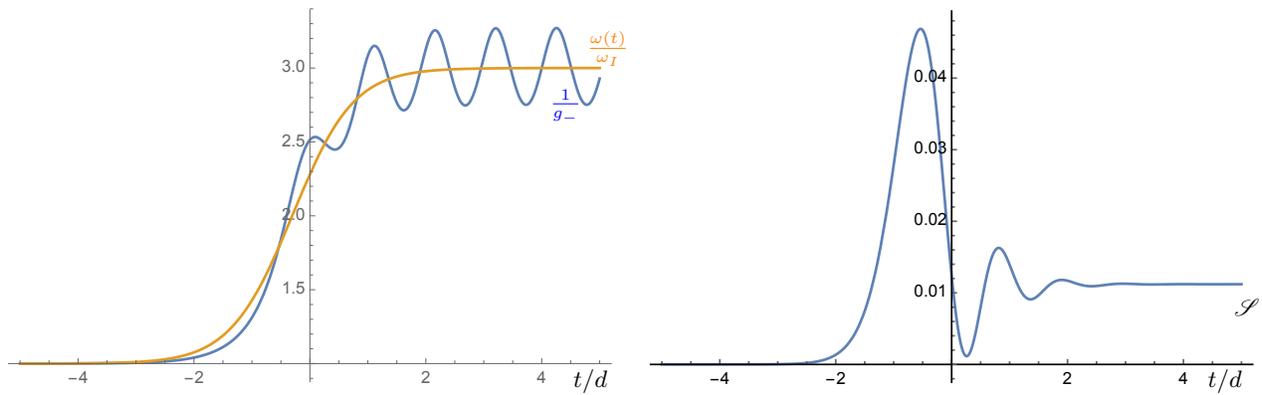
\begin{figure}[htb]
\begin{center}
\begin{tabular}{cc}
 \includegraphics[width=.45\linewidth,origin=tl]{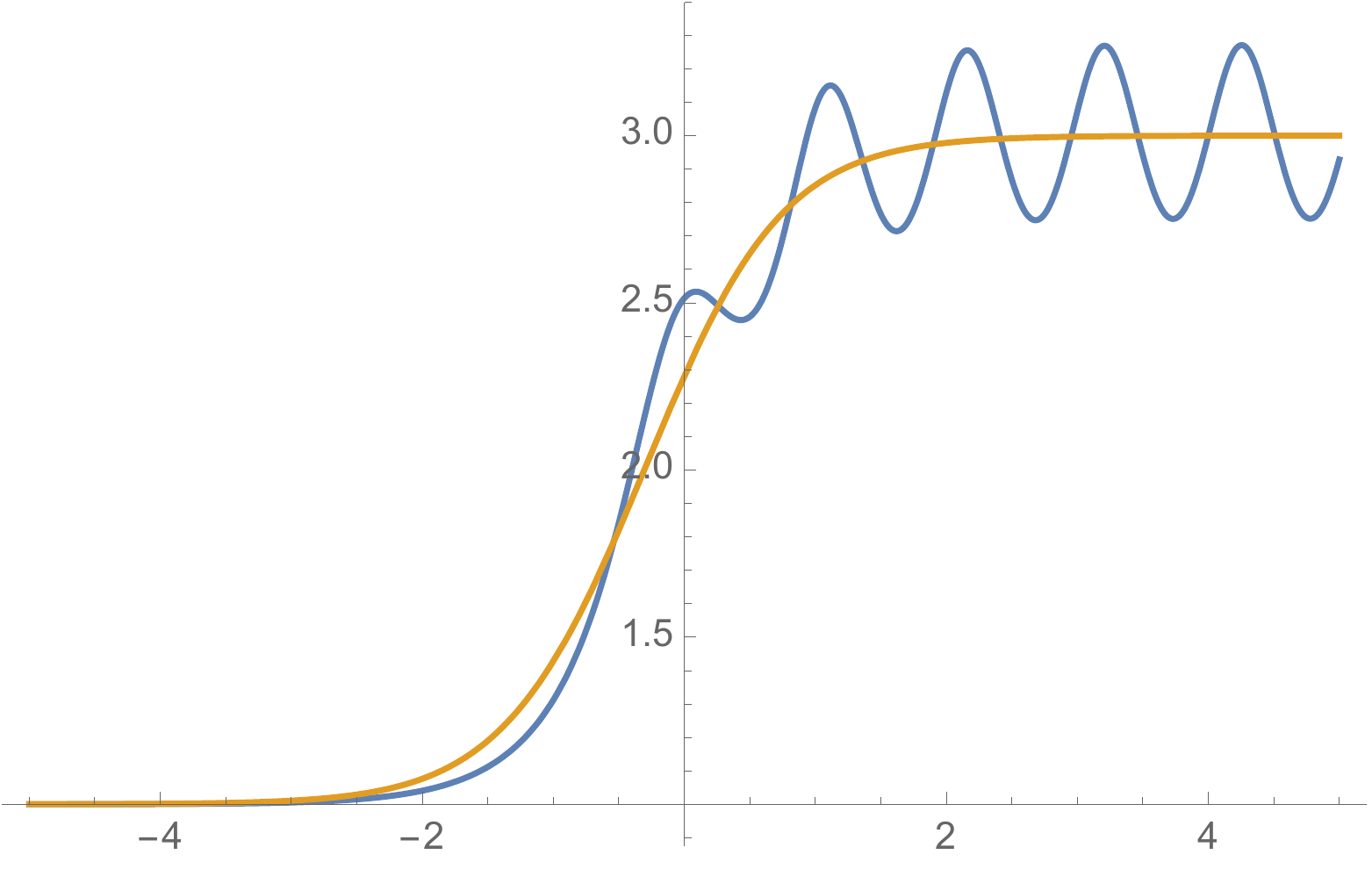} \qquad
 &~~\includegraphics[width=.45\linewidth,origin=tl]{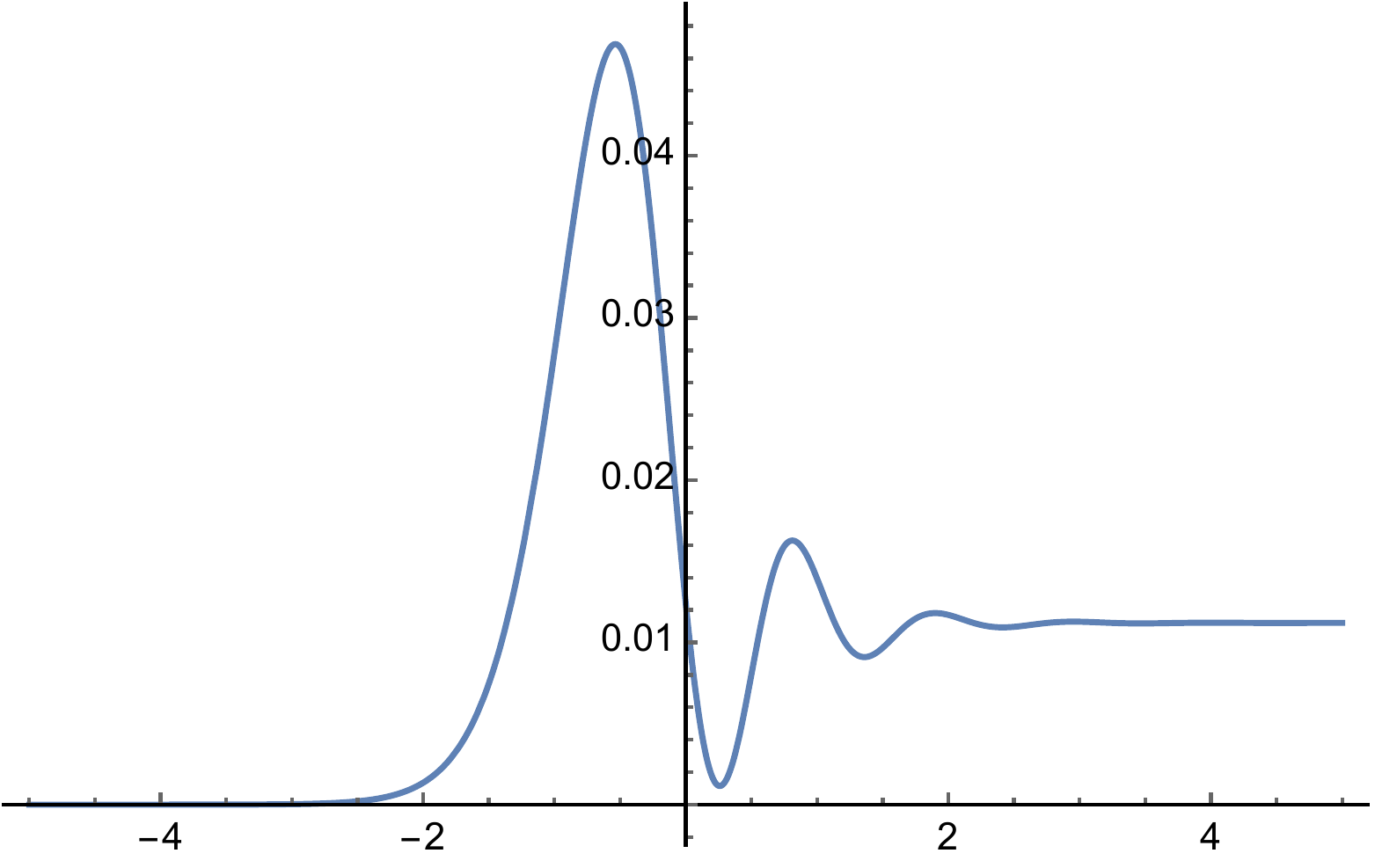} \\
\end{tabular}
\put (-20,-67) {$t/d$  }
\put (-260,-67) {$t/d$  }
\put (-269, 38) {\tcb{$ \frac{1}{g_-}$}}
\put (-10, -40) {$ \cS$}
\put(-255,59) {\textcolor{orange}{$\frac{\omega(t)}{\omega_I}$}}
\end{center}
\caption{A schematic form for the development of $g_-^{-1}$ and the $\cS$-factor $\cS$.
In this example, the frequency-squared follows the form in  Eq.~\eqref{omega t} and $\omega_0d = 5$, $\omega_-d = 1$, and $\omega_+/\omega_- = 3$, respectively.
}
\label{fig:non-a1}
\end{figure}

In the adiabatic and in the sudden jump limits, $\beta \to 0$ and $\beta \to (1- \omega_-/\omega_+)/2$, respectively.
The $g_-$ function becomes
$$
g_-(t) = 2m \omega_I g(t) g^*(t) = |_2F_1(\alpha_-, \alpha_+;1-i\omega_- d;y)|^2 .
$$
A characteristic form for $g_-^{-1}$ is plotted in Fig.~\ref{fig:non-a1}.
It oscillates around $\omega(t)/\omega_I$ with frequency $\omega(t)$.
As $t \to \infty$, it behaves as
$$
\lim_{t \to \infty} g_-(t) = |\alpha|^2+|\beta|^2+\alpha \beta^* e^{-2i \omega_+t} + \alpha^* \beta e^{2i \omega_+ t} .
$$
Now, we calculate the terminal value of the $\cS$-factor.
As $t \to \infty$, the $\cS$-factor becomes
\be{Sfactor}
\lim_{t \to \infty} \cS
= \lim_{t \to \infty} \frac{1}{2} \left[ \frac{\dot g_-^2}{4g_-}+ \frac{\omega^2}{\omega_I^2 g_-} \left(\frac{\omega_I}{\omega}- g_-\right)^2 \right]
 = 2\left(\frac{\omega_+ }{\omega_-}\right)^2 |\beta|^2.
\ee
Note that the value is closely related to the Bogoliubov coefficient $\beta$, which represents the change of the ground state.

Even though this formula for the $\cS$-factor is given only for the specific change~\eqref{omega t}, it contains various behaviors of the frequency.
We expect that this result holds at least qualitatively in other cases.
Explicitly,
\be{E: infty}
\lim_{t \to \infty} \cS =  \frac{\cosh [\pi(\omega_+-\omega_-)d]   +\cos [\pi \sqrt{1+(\varepsilon_+-\varepsilon_-)^2d^2}]}{2\sinh (\pi \omega_+ d) \sinh( \pi \omega_- d)} .
\ee
\begin{figure}[thb]
\begin{center}
\begin{tabular}{cl}
 \includegraphics[width=.4\linewidth,origin=tl]{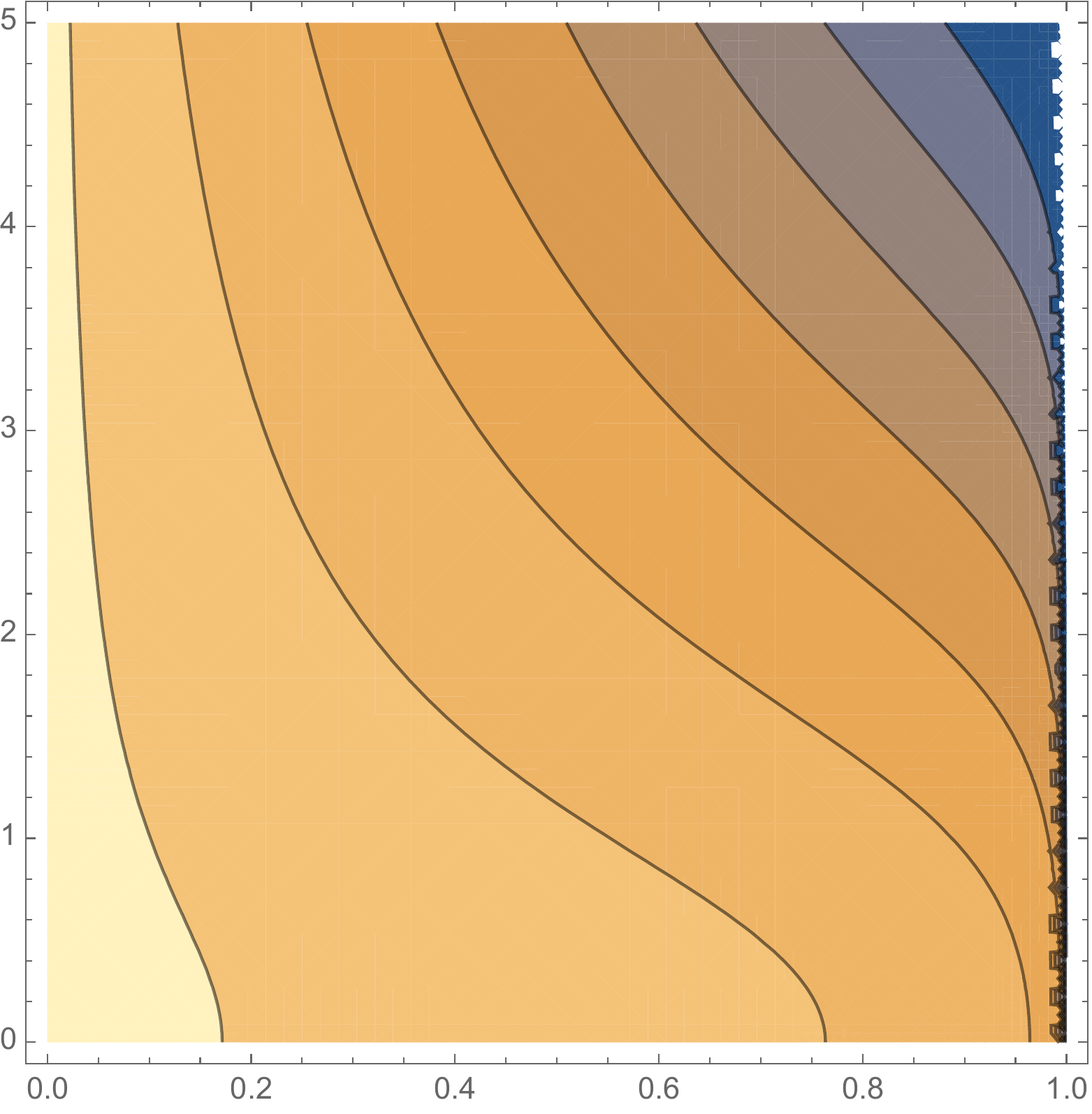}
 & \includegraphics[width=.05\linewidth,origin=tl]{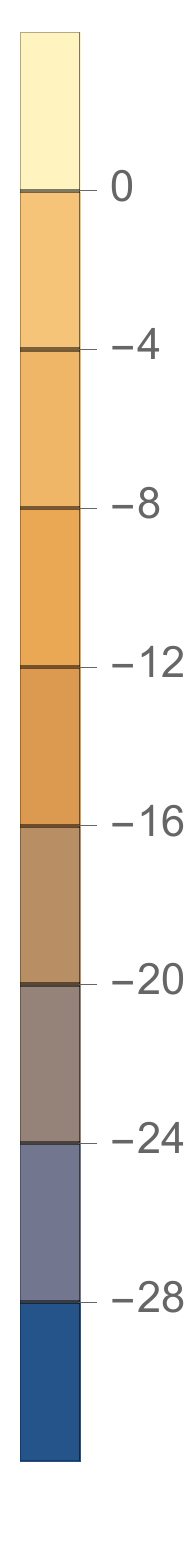} \\
\end{tabular}
\put (-65,-100) {$ \omega_+/\omega_-$  }
\put (-250, 78) {$ \omega_- d$}
\end{center}
\caption{Contour plot for the log of $\cS$-factor, $\displaystyle \lim_{t\to \infty}\log \cS$.
The numbers in the right column denote the value of $\log \cS$ on the line. 
Here, $\omega_0/\omega_- = 10$.
The $x$ and $y$ axes correspond to $\omega_+/\omega_-$ and $\omega_- d$, respectively.
The $\cS$-factor vanishes when $\omega_+=\omega_-$.
Because it is symmetric under $\omega_+ \leftrightarrow \omega_-$, we plot only the $0\leq \omega_+/\omega_-\leq 1$ region.
}
\label{fig:non-a2}
\end{figure}
Note that the developed $\cS$-factor is symmetric under the exchange $\omega_+ \leftrightarrow \omega_-$.
In the adiabatic limit $d\to \infty$, the $\cS$-factor exponentially decreases to zero, $e^{-2\pi \omega_- d}$.
In the sudden jump limit $d\to 0$, the $\cS$-factor becomes
\be{E: sudden jump}
\lim_{d\to 0,t \to \infty} \cS = \left[\frac12 \Big(\sqrt{\frac{\omega_+}{\omega_-}}
	-\sqrt{\frac{\omega_-}{\omega_+ }} \Big)\right]^2 .
\ee
Note that the value is finite and depends on the ratio between the two frequencies.
Therefore, only a finite amount of $\cS$-factor will be generated eventually however quickly the frequency changes.
There is other limit $\omega_+/\omega_-\to \infty$ keeping $\omega_-d$ finite.
In this limit, we have a thermal form with energy and temperature proportional to $\omega_-$ and $ 1/d$, respectively,
\be{E: large freq}
\lim_{\frac{\omega_+}{\omega_-}\to \infty, t\to\infty} \cS 
 = \frac1{e^{2\pi \omega_- d}-1}.
\ee
Now, we find that the produced $\cS$-factor diverges only after imposing both limits ($\omega_+/\omega_- \to \infty, ~d\to 0$).
We plot the behavior of the $\cS$-factor in Fig.~\ref{fig:non-a2}.

\section{ Thermodynamics of quantum harmonic oscillators undergoing nonadiabatic changes}\label{na-thermo}

Let us study the thermodynamics of a frequency-modulated quantum oscillator. 
We first deal with the time evolution of the initial thermal state in terms of the ELR invariant. 
Then, we study the general thermodynamics including squeezed states.

\subsection{Time evolution of the thermal state} \label{sec:IIIA}

The `invariant thermal state' defined in Eq.~\eqref{rho 0} is 
\be{density matrix b}
\rho_I = 2 \sinh \frac{\epsilon_I}2\, e^{-\hat I/T_0} 
=(1-e^{-\epsilon_I} ) 
\sum_{n=0}^{\infty}e^{-n \epsilon_I} |n \rangle_I  {}_I\langle n|,  \qquad
\epsilon_I\equiv \frac{\omega_I}{T_0}.
\ee

Let us consider the time evolution of an initial thermal state $\rho_0 = \rho_I$.
The energy $E=\langle H \rangle_I={\rm Tr}\left(\rho_I H(t)\right)$ for the `invariant thermal state' is, by using Eq.~\eqref{H t>0}, time dependent and has the value:
\bea
\label{EH}
E
&=&\frac{\omega}{4} \left(\frac{\omega_I}{g_- \omega} + \frac{\omega g_-}{\omega_I} + \frac{\dot g_-^2}{4\omega_I \omega g_-}\right) \coth \frac{\epsilon_I}{2}  \nn \\
&=& \frac{\omega_{\rm eff}}{2} \coth \frac{\epsilon_I}{2} ,  \label{H:I}
\eea
 where $\omega_{\rm eff}$ denotes
\be{omega eff}
\omega_{\rm eff} \equiv  \omega + \Omega, \qquad \Omega \equiv \omega_I\cS  .
\ee
Here we have used Eqs.~\eqref{eom E} and \eqref{E V} in the last equality.
As seen here, the energy is composed of two parts:
the original frequency part and the contribution of squeezing.
Note that from this energy dependency of the $\cS$-factor, 
we find that the $\cS$-factor is closely related to the $Q$-factor
of Husimi~\cite{10.1143/ptp/9.4.381} by $\cS = (Q-1)\omega/\omega_I$.

Let us obtain how the temperature changes in an adiabatic process.
Starting from the initial state $\rho_0$, we get the density matrix at $t$
\be{temp}
\hat\rho_0=\hat\rho_I=\rho_c e^{-\hat I/T_0} ~~~\rightarrow~~~\rho(t)=\rho_ce^{-\hat H(t)/T(t)}.
\ee
As we addressed in the introduction,
the reduced density matrix of a subsystem is described in terms of Gibbs distribution~\cite{Polkovnikov:2010yn,Calabrese:2011vdk}.
Until now, we calculate physical quantities in the Heisenberg picture.
The value of temperature at time $t$  should be independent on the picture.
To show the consistency, let us return to the Schr{\"o}dinger picture.
The invariant is approximated $I(t) \to g_-(t)H(t)$ from the relation \eqref{I:H g} in a quasi-static process.
For notational simplicity, let us write the density matrix at $t=t_0=0$ be given by $\rho_0$ for the time being.
Then, the density matrix at time $t=\delta t$, we have
\bea\label{T:dt}
\rho_{\rm Sch}(\delta t)&=&e^{-iH(\delta t)\delta t}\rho_ie^{iH(\delta t)\delta t}\nn\\
&=&e^{-iH(\delta t)\delta t}\rho_I(\delta t)e^{iH(\delta t)\delta t}\nn\\
&=&\rho_ce^{-iH(\delta t)\delta t}e^{-\frac{g_-(\delta t)}{T_0}H(\delta t)}e^{iH(\delta t)\delta t}\nn\\
&=&\rho_ce^{-g_-(\delta t)H(\delta t)/T_0}\nn\\
&=&\rho_I(\delta t) .
\eea
In a quasi-static (adiabatic) process all the way up from the initial time to the time $t$,
this continues, and we have
\be{T:t}
\hat \rho_{\rm Sch}(t)=\hat \rho_I (t)= 2 \sinh\left(\frac{\epsilon_I}{2}\right) e^{-\hat I/T_0} 
~=
2 \sinh \left(\frac{\epsilon_I}{2}\right)  e^{- \hat H(t)/T(t)}, \qquad
T(t) = \frac{T_0}{g_-(t)}.
\ee
The temperature of the oscillator is scaled by the $g_-^{-1}$ factor.
Therefore under the adiabatic process,
from Eq.~\eqref{ddot g-}, the temperature is written in terms of $\omega(t)$:
\be{Tw}
T(t) = \frac{T_0}{g_-(t)}=\frac{\omega(t)}{\omega_I}T_0.
\ee
We ensure this relation in Sec.~\ref{sec:IIIB}  by examining the first law of thermodynamics. 
If there is a nonadiabatic period in the middle of the process, 
this scaling of temperature no longer holds.
We obtain the temperature in this case later in Eq.~\eqref{Temperature}.

\subsection{ Nonadiabaticity}

The $\cS$-factor is motivated from a different perspective from the $Q$-factor discovered as a factor in the generating function of the transition probabilities.
Rather, $\cS$ here is closely related to $\langle W_{\rm irr}\rangle$, which was interpreted as `irreversible work' in Ref.~\cite{Galve2008}. 
In this article, we avoid the terminology because $\cS$ is closely related to the squeezing.

The time derivative of the $\cS$-factor~\eqref{non-adia} is
\be{nA}
\frac{d\cS}{dt} = \frac1{2\omega_I^2} \Big( g_-(t)- \frac{\omega_I}{\omega} \Big)\frac{d\omega^2(t)}{dt}.
\ee
Notice that this quantity becomes negligible during the time $\Delta t$ 
when
(1) $\omega^2$ is close to a constant, $\dot\omega\sim 0$,
(2)  the solution is around an extremum of the potential in Eq.~\eqref{E V}, $g_-(t)\sim\omega_I/\omega$.
As we discussed above, for the case (1), the process is slowly varying, i.e., adiabatic.
For the case (2), we also have $\dot\cE\sim 0$ but it does not vary slowly when $\dot\omega\neq 0$.
Even in this case,
the system can have an approximately conserved quantity for a short duration of time $\Delta t$.
Since $\Delta\left<W_{\rm irr}\right>=(\omega_I/2)\Delta\cE$,
$\Delta\cE\sim 0$ leads $\Delta\left<W_{\rm irr}\right>\sim 0$,
which means that the energy incremental during the time becomes free energy.
In addition, as we will show in the next subsection, the squeezing does not modify the temperature during the time.
Hence, even though the process is not quasi-static,
it can be approximated as
`adiabatic' 
around the local maximum or minimum of $\cE$ where $\dot\cE=0$
(see the right panel of Fig.~\ref{fig:non-a1}).
Consequently, the stiffness of the change in $\cS$-factor at a time $t$ is closely related to 
whether the process is adiabatic or not. 
In this sense,
we suggest 
\be{nA:E}
\nA\equiv \frac{d\cE}{dt}
\ee
 as a measure for `nonadiabaticity'.
With consideration of these, 
since $\cE$ is the integration of the nonadiabaticity $\nA$,
$\cE$ denotes a `stored' nonadiabaticity in the state.

Since the $Q$-factor is interpreted as ``the degree of adiabaticity'' in Ref.~\cite{Deffner_2008},
the nonadiabaticity is, in a certain degree, related to $dQ/dt$.
However,  with the above considerations,
we believe $\nA$ is more appropriate to represent  ``nonadiabaticity'' than $dQ/dt$.
With this reasoning and for the sake of practicality, 
we use $\cS$ rather than the $Q$-factor throughout this article.

\subsection{Thermodynamics under  quasi-static changes} \label{sec:IIIB}

Let us consider a thermal state  of the harmonic oscillator with a slowly varying $\omega(t)$ in the hamiltonian~\eqref{H:HO2}.
The thermal state $\hat \rho$ for the quantum harmonic oscillator is given by
\be{H rho}
\hat \rho \propto \exp\left[- \epsilon \Big(\hat a^\dagger \hat a+\frac12\Big)\right] ,
\ee
where $\hat a$ is a typical annihilation operator of the oscillator.
We have detached the subscript $(~){}_I$ under $\epsilon$ because it is not invariant anymore under the change of the entropy $S$ below. 
This density introduces a parameter $\epsilon(S)$ as a function of the entropy $S$ through
\be{S:ep}
S \equiv-\mbox{Tr}(\hat \rho\log \hat \rho) =  \frac{\epsilon}{e^{\epsilon}-1}
	-\log [1- e^{-\epsilon}],
\ee
which is a monotonically decreasing function from $\infty$ to zero as $\epsilon \in [0,\infty)$.

Therefore, the whole space of thermal states of harmonic oscillator is described by the three parameters $(m,\omega, S)$. 
The mass of the oscillator plays a characteristic role in distinguishing physical states in classical/quantum physics.
However, in thermodynamics of the harmonic oscillator, we are not interested in the individual form of the wave-function but interested in works and heat transfers which are related only to the energy change. 
As shown in Eq.~\eqref{H:I}, the energy of the system, $E\equiv \langle H\rangle =\frac{ \omega}{2} \coth \frac{\epsilon}{2}$ is independent of the mass for the present thermal state.
Consequently, it is enough for us to consider a set of all thermal states of different frequency $\omega$ and of different entropy $S$.
Thus, we represent the space of all thermal state with a set of non-negative numbers, $(\omega,S)$.

Considering two nearby systems by $\delta S$ and $\delta \omega$, the energy difference will be related by the first law
\be{1st:adiabatic}
\delta E = -F_\omega \delta \omega+T \delta S ,
\ee
where the force term and the temperature can be written as
\be{F:adia}
F_\omega(S)  \equiv -\left(\frac{\partial E}{\partial \omega}\right)_{S}
	=-\frac12 \coth\frac{\epsilon(S)}{2}, \qquad
T(\omega,S)  \equiv \left(\frac{\partial E}{\partial S}\right)_{\omega}  =
\omega\frac{d\epsilon}{d S}\frac{d }{d\epsilon}\left(\coth\frac{\epsilon}{2}\right)= \frac{\omega}{\epsilon(S)},
\ee
from Eq.\eqref{EH}. 
Here, the subscripts $S$ and $\omega$ denote that their values are held.
The temperature is consistent with the previous result~\eqref{Tw} obtained from the adiabatic (quasi-static) invariant.
For later convenience, we present a variational formula for the entropy~\eqref{S:ep}: 
\be{delta S:delta ep}
\delta S = - \frac{\epsilon}{4 \sinh^{2} (\epsilon/2)} \delta \epsilon .
\ee

\subsection{Thermodynamics in the presence of nonadiabaticity} \label{sec:non-adia}

In this subsection, we argue that the $\cS$-factor plays the role of a thermodynamic quantity similar to the temperature, the entropy and the energy.

A nonadiabatic thermal state $\hat \rho$ for a quantum harmonic oscillator is given by
\be{H rho}
\hat \rho \propto \exp\left[- \epsilon \left(\hat b^\dagger \hat b+\frac12\right)\right] ,
\ee
where the thermal state is created over the vacuum which is annihilated by the operator $\hat b$ in Eq.~\eqref{invariant op}.
The thermal state depends on  parameters $m$, $\omega$ in the hamiltonian, $\epsilon$ in the density, and $(\omega g_-/\omega_I)$, $\dot g_-/\omega_I$ in the operator $\hat b$.
Fortunately, the mass $m$ does not affect on the thermodynamics and
 the parameters $(\omega g_-/\omega_I)$ and $\dot g_-/\omega_I$ affect on 
 the energy only through their combination, $\omega_I\cS$.
Consequently, the parameter space of thermal states is described by the three parameters, $(\omega, \epsilon, \omega_I\cS)$.

Now, let us consider two nearby systems by the small variations $\delta \omega$, $\delta S$, and $\delta \cS$, where the entropy variation $\delta S$ is related to $\delta \epsilon$ through Eq.~\eqref{delta S:delta ep}.
The first law of thermodynamics can be derived by varying the energy expectation value in Eq.~\eqref{H:I} with respect to the general variations to get
\be{1st law}
\delta E(\omega_{\rm eff},S) =-F_\omega \delta \omega_{\rm eff}+ T\delta S,
\ee
where $\omega_{\rm eff}$ is an effective frequency in Eq.~\eqref{omega eff}. 
The force term takes the same form as that of the adiabatic case
\be{Force}
F_\omega(S) \equiv - \left(\frac{\partial E}{\partial \omega_{\rm eff}}\right)_{S}
	=-\frac12 \coth \frac{\epsilon}{2}.
\ee	
This force term is the same as the work term of the classical oscillator in Ref.~\cite{Boyer:2019irq}.
One can vary the energy with respect to the entropy independently to find the oscillator's temperature from the thermodynamic relation,
\be{Temperature}
T(\omega_{\rm eff},S)  \equiv \left(\frac{\partial E}{\partial S}\right)_{\omega_{\rm eff}}
=  \frac{\omega_{\rm eff}} {\epsilon}  = \frac{ \omega + \Omega}{\epsilon} .
\ee

In the limit $\cS \to 0$, Eq.~\eqref{Temperature} reproduces the adiabatic result~\eqref{Tw}.
If there are nonadiabatic durations in the process, 
the effects make a difference from the adiabatic result and
the information is stored into $\cS$.
Due to the contribution from the nonadiabatic period, the temperature is always higher than the adiabatic value because $\cS \geq 0$.
From these observations, the first law of thermodynamics can be rewritten to include 
the net effect of nonadiabatic evolution at the present harmonic oscillator system.

\subsection{Physical constraints }
The variation of $\omega_{\rm eff}$ is composed of two parts:
$$
\delta \omega_{\rm eff} = \delta \omega + \delta \Omega .
$$
In general, the two variations $\delta \omega$ and $\delta \Omega$ are independent.
However, for a specific thermodynamic process, the variations must be related by physical constraints.
For example, $\delta\Omega=\omega_I\delta\cE$ in \eqref{eom E} is determined by 
the frequency-modulated force and the inverse-cubic force in \eqref{eomh}.
Therefore, to reproduce the evolution of the quantum harmonic oscillator, 
we should relate the two variations $\delta\omega$ and $\delta\Omega$ properly.
As an example, let the frequency variation $\delta \omega$ is related to the variation $\delta \Omega$ by 
\be{dnA:domega}
\Omega' \delta \omega = \delta \Omega. 
\ee
This implies that $\Omega$ is just a function of $\omega$ only.
Then, the nonadiabaticity is automatically determined from the change of the frequency $\omega$ by
\be{dA dt}
\omega_I \nA = \frac{d\Omega}{dt} = \Omega' \dot \omega.
\ee
The effective frequency is determined to be $
	\delta \omega_{\rm eff} = ( 1+ \Omega') \delta \omega .$
Conversely, given $\Omega'$ at an instance, we can determine the frequency change $\delta \omega$ from $\delta \Omega$. 
If we consider only monotonic changes of frequency, the one-to-one correspondence between the time and $\omega$ justifies it.
This relation~\eqref{dnA:domega} plays the role of an equation of state for the oscillator thermal system in the sense that it determines the temporal evolution.
More specifically, the frequency-modulated oscillator in Sec.~\ref{example} can be reproduced 
if we choose $\Omega'$ to satisfy
\be{omega gm}
	\frac{\omega g_-}{\omega_I}
		= 1+ \Omega',
\ee
where we use Eq.~\eqref{nA}.
Then, $\Omega$ and $\Omega'$ determine $\dot g_-/\omega_I$ to be
$$
\frac18 \left( \frac{\dot g_-}{\omega_I}\right)^2 = \left(\frac{\omega g_-}{\omega_I}\right) \left( \frac{\Omega}{\omega}\right)
 	- \frac12 \Big(1 -\frac{\omega g_-}{\omega_I} \Big)^2  
= \left(1+\Omega' \right) \frac{\Omega}{\omega}
 	- \frac{\Omega'^2}{2}.
$$ 
As seen here, specifying $\Omega$ and $\Omega'$ fully determines the nonadiabatic evolution of an initial thermal state of the oscillator. 

\begin{figure}[bth]
\begin{center}
\begin{tabular}{cc}
 \includegraphics[width=.5\linewidth,origin=tl]{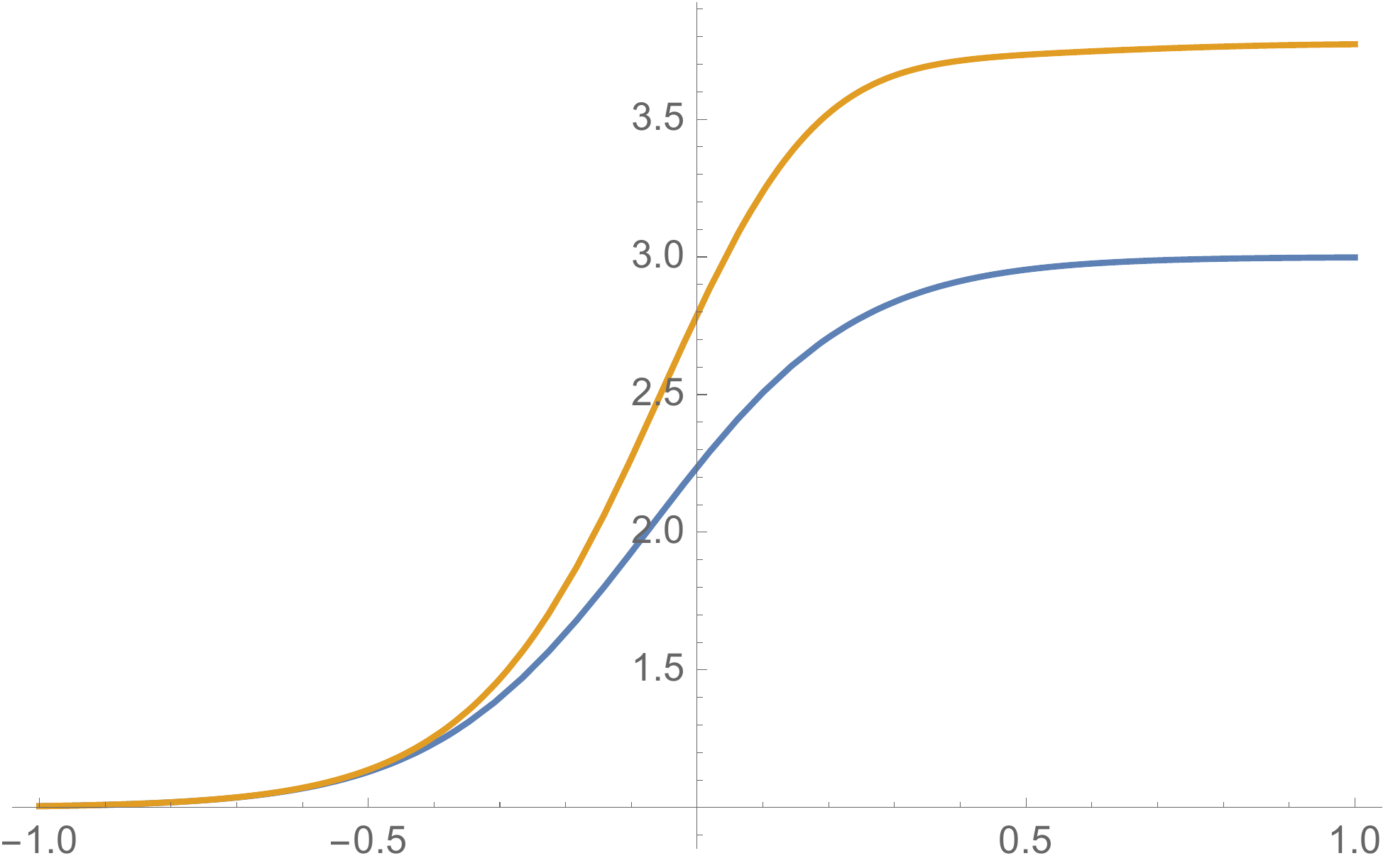}\\
\end{tabular}
\put (-60,38) {$\omega(t)/\omega_I$  }
\put (-70,79) {$ \frac{T(t)}{T_0}$  }
\put (-10,-60) {$\omega_I t$}
\end{center}
\caption{
Evolution of the temperature for the  frequency in Eq.~\eqref{omega t}.
Here we take $\omega_+/\omega_- = 3$, $\omega_- d= 0.3$, and $\omega_0^2=\omega_+^2+(40\omega_-)^2$.
}
\label{fig:T}
\end{figure}

As an exercise, let us reconstruct the evolution of the oscillator given in Fig.~\ref{fig:non-a1}.
Given $\Omega$ and $\Omega'$, one may obtain $\omega g_-/\omega_I$ and $\dot g_-/\omega_I$ from the above two equations. 
Then, the thermal state is determined because the annihilation operator is given by Eq.~\eqref{invariant op} once the entropy is given.
Given $\Omega$ at the next instance of time, we get $d\Omega/dt $ and then we get $\dot \omega$ from Eq.~\eqref{dA dt}. 
Therefore, arranging appropriate values of $\Omega$ at each instance of time determines the time dependence of the frequency $\omega$. 
For example, one may arrange $\Omega = \omega_I \cS$ so that $\cS$ follows the curve in the right panel of Fig.~\ref{fig:non-a1}.
Then, one naturally gets the frequency $\omega$ given by Eq.~\eqref{omega t}. 
In Fig.~\ref{fig:T},
we plot the temperature of the oscillator in Eq.~\eqref{Temperature}.

\section{ Summary and discussions}\label{summary}

In this work, we have considered a (non-)adiabatic time evolution of a frequency-modulated quantum harmonic oscillator.
We introduced a general method to quantity for `nonadiabaticity' of a system undergoing thermodynamic changes.
We noted that the ELR invariant plays a central role.
We have also studied the thermodynamics of the oscillator  
focusing especially  on its first law.

We use the ELR invariant state in place of the initial state to analyze nonadiabatic processes. 
For a time varying hamiltonian $\hat H(t)$, the ELR invariant $\hat I$ plays a key role in describing the evolution of quantum state.
For a quasi-static (adiabatic) process,
the invariant is nothing but a scaled hamiltonian $g_-(t)H(t)$, where $g_-(t)$ is a slowly varying function.
Hence, nonadiabaticity appears when the difference between $\hat I$ and $g_-\hat H$ does not vanish.
For a general nonadiabatic process, we have obtained a dimensionless quantity $\cS$, 
which represents a stored nonadiabaticity during the process.
We found that this quantity $\cS$ is always non-negative and is closely related to the energy used in squeezing the ground state.

We showed that the $\cS$-factor here converts to the $Q$-factor of Husimi,
and it can also be related to the so-called `irreversible work' 
in Ref.~\cite{Galve2008}, the squeezing parameter~\cite{Kim:2003mi}, and the Bogoliubov coefficient $\beta$.
From a quite different perspective,
we could satisfactorily quantify `nonadiabaticity' as $\nA\equiv d\cE/dt$
and show its  advantages over $dQ/dt$.
Literature have suggested `nonadiabaticity' in terms of various quantities~\cite{Yonehara2012,Zimmermann2010,Zimmermann2012}.
In these references, the authors proposed measures for `nonadiabaticity' suitable for 
the molecular quantum dynamics in some situations.
In this work, we suggest a measure for `nonadiabaticity' for a quantum oscillator.

Let's ask whether we can apply the prescription for getting the nonadiabaticity
to other systems than the harmonic oscillator. 
A different  hamiltonian $\tilde H(t)$ (not the oscillator one in \eqref{H:HO2}) will present another invariant $\tilde I(t)$.
As we saw  in Sec.~\ref{review}, a set of temporal functions will determine $\tilde I(t)$.
The invariant condition for $\tilde I$ determines a set of differential equations for
the functions.
As we see in Eq.~\eqref{eomh}, one can construct a similar energy-like time-dependent quantity $\tilde\cE$.
It is not certain that $\tilde\cE$ still satisfy the non-negativeness.
It is because the positivity of the net energy flow for the equation is not guaranteed in general.
However, we think the quantity $\tilde\nA = d\tilde\cE/dt$  still can be interpreted as `nonadiabaticity'.
It is because the condition $d\tilde\cE/dt=0$ leads that the temporal parameter of the system, e.g., $\dot \omega(t)$ in Eq.~\eqref{nA}, is slowly varying.
So, for a duration of time when $d\tilde\cE/dt=0$, the process can be interpreted as `adiabatic'.
Thus, we interpret the condition $d\tilde\cE/dt\neq 0$ as `nonadiabatic' naturally.

In this work, we have extended the parameter space of the thermal state from $(\omega, S )$ for Gaussian thermal states to $(\omega, S, \cS)$ for squeezed-Gaussian states.
Given a temporal frequency change for an initial thermal state $(\omega_I, S, 0)$ of the quantum oscillator, we have shown the followings:
If the change is quasi-static, the temperature of the oscillator is proportional to the frequency, $T(t) = (\omega(t)/\omega_I)T_0$. For non-quasi-static changes, a more general form of the temperature, $T(\omega_{\rm eff},S)  =  ( \omega + \omega_I\cS)/\epsilon$, was given.
Here, we introduced an effective frequency $\omega_{\rm eff} = \omega + \Omega$, 
where $\Omega(\equiv \omega_I\cS)$ denotes the nonadiabaticity contribution.
We wrote the first law of thermodynamics similar to the typical form, $dE = -F_\omega d\omega_{\rm eff} + T dS$.
Intriguingly, the squeezed states obey the same first law as an adiabatic thermal state with modified temperature $T$ and frequency $\omega_{\rm eff}$.

In Ref.~\cite{Pechal}, the authors studied the measurability and controllability of nonadiabatic effects of an electronic harmonic oscillator to the geometric phase. They discussed nonadiabatic contribution $\Delta_{na}$ to the geometric phase in addition to the adiabatic contribution, the difference of geometric phases for two nearby closed cycles is given by $\delta\phi =\delta\phi_a+\Delta_{na}$. 
Thus, one can infer that the  nonadiabaticity contribution to the first law is directly related to that of  $\Delta_{na}$ to the geometric phase.
Further, one can obtain the direct relation as $\Delta_{na}=\int^B_A V_{I}(F_\omega)d\omega_I$ for a process from a state $A$ to $B$. Obtaining an exact form of $V_{I}(F_\omega)$ is an interesting study.
In light of this consideration, at least for a harmonic oscillator system,
one can conclude that the nonadiabatic contribution to the first law is measurable.

One may also use the method developed here for the thermodynamic analysis of various systems.
One example is applying it for thermodynamics of the charged particle systems in Ref.~\cite{PhysRevLett.18.510} moving in a classical, axially symmetric uniform magnetic field.
When the magnetic field increases and becomes extremely strong, 
this system is known to manifest noncommutativity.
Hence, a similar analysis of this study would hint at the thermodynamics in a noncommutative space in a natural way.

\section*{Acknowledgment}
This work was supported by the National Research Foundation of Korea grants funded by the Korea government NRF-2020R1A2C1009313.


\end{document}